\documentclass[10pt]{article}    
\usepackage[margin=0.75in]{geometry}
\usepackage{setspace}
\usepackage{helvet}

\usepackage{amsthm}
\usepackage{booktabs}
\usepackage{graphicx, color}
\usepackage[sort&compress,numbers,square]{natbib}
\usepackage{longtable}
\usepackage{afterpage}  
\usepackage{hyperref}
\usepackage{soul}
\usepackage{titlesec}
\usepackage{subcaption}

\titlespacing{\section}{3pt}{3pt}{3pt}
\titlespacing{\subsection}{1pt}{*1}{*1}
\titlespacing{\subsubsection}{1pt}{*1}{*1}

\title{The Elephant in the Room: Software and Hardware Security Vulnerabilities of Portable Sequencing Devices}
\author{
    Carson Stillman$^1$
    \and Jonathan E. Bravo$^1$ 
    \and Christina Boucher$^1$
    \and Sara Rampazzi$^{1,\star}$
}

\date{
	$^1$Department of Computer and Information Science and Engineering,\\
 Herbert Wertheim College of Engineering, 
    University of Florida \\ 
    $^{\star}$Corresponding author: srampazzi@ufl.edu}

\begin{document}

\maketitle


  \vspace{-8mm}
\section*{INTRODUCTION}
\label{sec:intro}
\begin{figure}[h]
   \centering
   \includegraphics[width=0.70\textwidth]{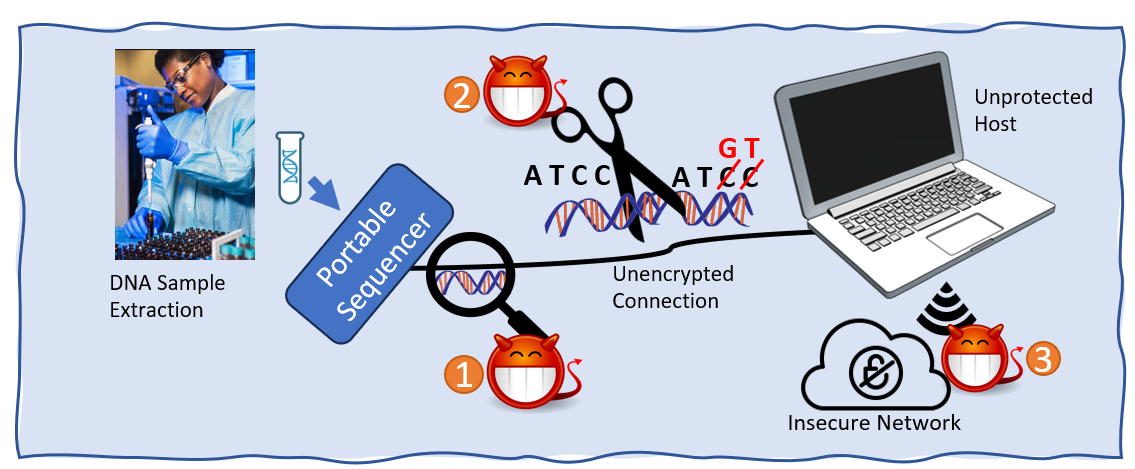}
  \vspace{-2mm}
   \caption{{\small\textsuperscript{} Potential Adversarial Attack Paths in portable sequencing devices. (1) Without encryption, adversaries have the ability to read the signal as it is sent via cable to external host computers. (2) Without integrity checks, adversaries can modify the signal to produce misclassification of called bases. (3) Without robust access control, adversaries can access sequenced information and start/stop collections while physically far away from the sequencing device.}}
   \label{fig:main}
   \vspace{-5mm}
\end{figure}

\noindent Portable genome sequencing technology is revolutionizing genomic research by providing a faster, more flexible method of sequencing DNA and RNA~\cite{nanopore_improvements,nanopore_smaller}. The unprecedented shift from bulky stand-alone benchtop equipment confined in a laboratory setting to small portable devices which can be easily carried anywhere outside the laboratory network and connected to untrusted external computers to perform sequencing raises new security and privacy threats not considered before. 
Current research primarily addresses the privacy of DNA/RNA data in online databases~\cite{database_sec} and the security of stand-alone sequencing devices such as Illumina~\cite{alsaffar2022digital}. However, it overlooks the security risks arising from compromises of computer devices directly connected to portable sequencers as illustrated in Fig.~\ref{fig:main}.
While highly sensitive data, such as the human genome, has become easier to sequence, the networks connecting to these smaller devices and the hardware running basecalling can no longer implicitly be trusted, and doing so can deteriorate the confidentiality and integrity of the genomic data being processed. 

Here, we present new security and privacy threats of portable sequencing technology and recommendations to aid in ensuring sequencing data is kept private and secure. First, to prevent unauthorized access to sequencing devices, IP addresses should not be considered a sufficient authentication mechanism. Second, integrity checks are necessary for all data passed from the sequencer to external computers to avoid data manipulation. Finally, encryption should be considered as data is passed from the sequencer to such external computers to prevent eavesdropping on data as it is sent and stored. As devices and technology rapidly change, it becomes paramount to reevaluate security requirements alongside them or risk leaving some of our most sensitive data exposed.  

\section*{ANALYSIS AND OBSERVATIONS}
\label{sec:results}

\noindent{\bf Confidentiality.} Confidentiality is the security property that guarantees that unauthorized parties are not able to view and extract sensitive information. 
This is especially critical in genomics research as samples may be sourced from human participants and compromised devices could lead to the exposure of personal identifiable information, potentially violating laws such as HIPAA. 

Portable devices typically delegate their computational tasks to external host systems, such as laptops or tablets~\cite{minion_it_requirements}. This process necessitates the transfer of data between the portable device and the host system during computation. Additionally, the device may store output results on the host computer through its application software. 

As corporate policies like Bring Your Own Device (BYOD) become more prevalent, employees and end users are increasingly using unprotected personal devices on company networks. These devices are used to access corporate data and perform job duties remotely, increasing the likelihood of having a compromised computer connected to a portable sequencer.
For instance, studies have shown that adversaries, using automated tools like Wireshark~\cite{wireshark} or malware programs~\cite{ney2017computer} subtly installed on a compromised computer, can monitor and extract information of sequencing tasks. This exposes confidential data and potentially allows data transmission to unauthorized third parties. It has thus become imperative to safeguard data in such a way that it becomes unreadable to potential adversaries, for instance using encryption algorithms to encrypt data passed from the sequencer to external hosts.

\noindent\textbf{Authentication.} Portable sequencing devices might also connect to external computers connected to the internet to allow for remote monitoring of sequencing execution. 
Strong authentication mechanisms should be in place to protect data confidentiality in such scenarios. 
For instance, IP addresses should not be used as the only information required to authenticate a remote user for interaction with a portable sequencer. 
Several studies have shown how IP addresses can be found by malicious adversaries on the same network, including on restricted networks and Virtual Private Networks (VPNs), via a simple port scan~\cite{nmap}.

Furthermore, the risk to the sequencing device and its data is significantly increased when host systems are connected to unsecured public networks outside the corporate network. This situation often arises during fieldwork in commercial agricultural settings or within clinical environments.
To ensure the security of sequencing devices against unauthorized remote access, it is essential to implement passwords and similar strong authentication protocols. Relying solely on VPNs and private networks does not provide sufficient protection nor assure security. 
\vspace{-0.5cm}

\medskip


\noindent{\bf Integrity.} Integrity is a security property that ensures data transmitted between two parties arrives to the recipient as intended, guaranteeing it has not been tampered with. This protects researchers and industry professionals from an adversary altering data as it is transmitted or received, such as changing sequencing outcomes or falsifying results. As research teams have already demonstrated how DNA can be altered on benchtop sequencing equipment~\cite{ney2017computer}, such threat escalates for portable devices if connected to compromised host computers.

For instance, to mitigate such risk the use of Message Authentication Codes (MACs) on sequencing data can reveal tampering during data sharing. These codes are calculated first by the sender, who performs mathematical operations on the data which results in a close-to-unique value, that is attached to the end of the data as it is transmitted. Upon receipt of the data, the receiver performs the same mathematical operations to compare their calculated value. If this matches the value sent, the message has not been tampered with. 



\section*{CONCLUSION}
\label{sec:conclusion}
\noindent To mitigate vulnerabilities of portable devices, a multi-layered security approach is essential. This includes regular software updates, strong data encryption, comprehensive access control, regular security assessments, and user education on security best practices. Given the critical nature of the data processed by sequencing platforms, revisiting security and privacy requirements for portable hardware and software is crucial to protect against data breaches, ensure data integrity, and maintain trust in biological research and clinical diagnostics before sequencing data are stored in online databases. 

\footnotesize
\bibliographystyle{ieeetr} 
\bibliography{output.bbl}

\end{document}